\documentclass[11pt,notitlepage]{article}
\usepackage{graphicx}
\usepackage[english]{babel}
\usepackage{array}
\usepackage{placeins}
\usepackage{tabu}
\usepackage{float}
\usepackage{amsmath} 
\usepackage{mathrsfs}
\usepackage{physics}
\usepackage{hyperref}
\hypersetup{colorlinks,linkcolor={blue},citecolor={blue},urlcolor={red}} 

\usepackage[sort&compress]{natbib}
\usepackage{amsfonts}
\usepackage{amssymb}
\usepackage[left=30mm,right=20mm,top=30mm,bottom=22mm]{geometry}
\usepackage{multicol}
\usepackage{url}

\usepackage[immediate]{silence}
\usepackage{sectsty}
\usepackage{lipsum}
\sectionfont{\fontsize{14}{14}\selectfont}
\subsectionfont{\fontsize{12}{14}\selectfont}
\chapterfont{\centering}
\usepackage{tikz}

\date{}

\providecommand{\keywords}[1]
{
  \small	
  \textbf{\textit{Keywords---}} #1
}

\title{Entanglement Evolution of Noisy Quantum Systems: Master Equation-TFD Solutions}
\author{Urjjarani Patel and KVS Shiv Chaitanya \\
        \small Department of Physics, BITS-PILANI Hyderabad campus, Hyderabad, 500078, India    }
\date{} 

\begin{document}
\maketitle
\pagenumbering{arabic}

\begin{abstract}
In this paper, Thermofield Dynamics (TFD) is applied to map a quantum optics nonlinear master equation into a Schrödinger-like equation for any arbitrary initial condition. This formalism provides a more efficient way for solving open quantum system problems. Then we use the Hartree-Fock approximation to solve the master equations of two separate noisy quantum systems analytically, which allows us to analyze the entanglement and quantum mutual information in each case using the eigenvalues of a covariance matrix, followed by two-mode and single-mode squeezed states.  

\end{abstract} \hspace{10pt}



\section{Introduction}
The unitary time evolution of pure quantum states is characterized by the Schrödinger equation under isolated conditions and governs the evolution of closed quantum systems \cite{nielsen2010quantum}. However, the standard Schrödinger formalism cannot solve all the realistic physical systems, which invariably interact with their surrounding environment, leading to non-unitary dynamics. Mostly, when a quantum system of interest is coupled to 
its surrounding environment, it leads to the evolution of dissipative behavior, explained by the master equation. Thus, the quantum master equation provides the theoretical framework for describing such open quantum systems, decoherence, and entanglement, which implies the time evolution of the system's density matrix while accounting for the influence of environmental interactions \cite{chaitanya2011algebraic}.

In quantum mechanics, one of the ways entanglement arises is when a system interacts with its surrounding environment, leading to an inseparable state due to quantum fluctuation \cite{hashizume2013understanding}. Many challenges in entanglements, including bipartite or multipartite systems \cite{hillery2010conditions}, attract researchers to resolve them in various approaches. The Peres criterion \cite{peres1996separability} provides a simple way of calculating the entanglement of a system, called negativity. This can be calculated by taking the negative eigenvalue of the transpose of a covariance matrix, and the logarithmic negativity gives researchers a suitable upper bound to distillable entanglement \cite{boura2016dynamics}.  

The master equation originated historically in quantum optics and serves as a theoretical framework for understanding fundamental phenomena, which include spontaneous emission \cite{agarwal1970master}, optical cavity \cite{barlow2015master}, and thermal equilibration processes \cite{alexander2011approach}. Apart from the foundational role in quantum optics, the master equation approach has proven invaluable for modeling decoherence mechanisms that limit quantum coherence, analyzing the dynamics of quantum entanglement, and characterizing the performance of quantum information processing protocols in noisy environments. The mathematical solutions of the master equation provide direct access to crucial quantum information measures such as concurrence, logarithmic negativity, channel capacities, and other entanglement monotones that quantify the quantum correlations available for information processing tasks, which makes the master equation framework essential for both theoretical analysis and practical implementation of quantum technologies \cite{chaitanya2011algebraic}.

This formalism provides significant applications in quantum optics as well as in condensed-matter systems. It captures the realistic experimental situations where environmental interactions, decoherence, and dissipative mechanisms cannot be ignored \cite{cubitt2012complexity}.

To tackle the challenges of solving master equations, several approaches have been taken in quantum information theory. The most popular method is the Glauber-Sudarshan P-representation, which cleverly transforms the quantum master equation into a classical-like c-number Fokker-Planck equation that is much easier to handle mathematically \cite{breuer2002theory,gardiner2004quantum}. Also, the series method has proven effective when dealing with pair coherent states \cite{agarwal1986generation}. At the same time, the usage of thermal coherent states has bridged the gap between master equations and Fokker-Planck equations \cite{fan2002new}. Subsequently, other researchers have employed the sophisticated technique of integration within an ordered product (IWOP) of operators \cite{fan2006newton} to explicitly calculate the Kraus operators that explain how Kerr media affect quantum systems \cite{hu2009infinite}.

Meanwhile, the thermo field dynamics (TFD) offers a robust framework for solving master equations in Kerr media by exploiting the disentanglement theorem, regardless of what initial conditions you start with \cite{chaitanya2011algebraic,shanta1996eigenstates,shanta1996operator}. It differs from other methods by doubling the Hilbert space. TFD is utilized to simplify quantum logic in the development of quantum circuits \cite{lanyon2009simplifying,miceli2019thermo}. On the other hand, the TFD approach presents an advantage in investigating gate-based quantum computers because it is a real-time operator-based approach that uses Bogoliubov transformations of quantum field theory at finite temperature \cite{sagastizabal2021variational}. Implementation of qubits as well as logic gates at finite temperature using the TFD technique provides many challenges for researchers \cite{lee2022variational,miceli2019thermo,sagastizabal2021variational,prudencio2019quantum,rowlands2018noisy,wu2019variational}. Also, this technique is used to describe quantum electron vibrational dynamics in molecular systems at finite temperatures. The simulations reveal the impact of certain vibrational modes on exciton dynamics and energy transfer, emphasizing the need for accurate modeling of electron-phonon coupling \cite{borrelli2017simulation}. A detailed explanation of the TFD is given below.

\subsection{Introduction to TFD}
TFD assists in the conversion of the master equation to a Schrödinger-like equation, and the methods available for solving the Schrödinger equation can be applicable for further calculations. In TFD formalism, a density operator $\rho=\vert N\rangle\langle N\vert$ corresponding to a Fock state $\vert N\rangle$ in the Hilbert space ${\cal H}$ is viewed in TFD as a vector $\rho=\vert N, \tilde{N} \rangle$ in an extended Hilbert space ${\cal H}\otimes {\cal H}^*$. The central idea in TFD is to construct a density operator $\vert \rho^\alpha\rangle,  1/2  \le \alpha \le 1$ as a vector in the extended Hilbert space ${\cal H}\otimes {\cal H}^*$. A brief description of TFD is given in Appendix A.\\
Here, the averages of operators with respect to $\rho$ reduce to a scalar product:
\begin{equation}
\langle A\rangle = Tr [A\rho] = \langle\rho^{1-\alpha}\vert A\vert \rho^\alpha\rangle,
\end{equation}
where $\vert \rho^\alpha\rangle$ is given by
\begin{equation}
\vert \rho^\alpha\rangle = \hat{\rho}^\alpha\vert I\rangle\;, \text{with},\; \hat{\rho}^\alpha = \rho^\alpha \otimes I,
\end{equation}
where $\vert I\rangle$ is the resolution of the
identity
\begin{equation}
\vert I\rangle=\sum \vert n\rangle\langle n\vert = \sum \vert n\rangle\otimes\vert \tilde{n}\rangle \equiv \sum \vert n,\tilde{n}\rangle,
\end{equation}
in terms of a complete orthonormal basis  $\{\vert n\rangle\}_{n=0}^\infty$  in  ${\cal  H}$.
The state vector $\vert I\rangle$ takes a normalized vector to another
normalized vector in the extended Hilbert space ${\cal H}\otimes {\cal 
H}^*$. The matrix $A(a,a^\dagger)$  acts like $A \otimes I$. 

(It may be noted that for any density operator the states $\vert \rho^\alpha\rangle,  1/2  \le \alpha \le 1$ have a finite norm in the extended Hilbert space ${\cal H}\otimes {\cal H}^*$. This is not in general true for the state $\vert \rho^{1-\alpha}\rangle,  1/2  \le \alpha \le 1$, which includes $\vert I\rangle$. These states are regarded as formal but extremely useful constructs.)\\
The creation and the annihilation operators $a^\dagger, \tilde{a}^\dagger,  a$,  and  $\tilde{a}$ are introduced as follows
\begin{equation}
\begin{split}
a\vert n,\tilde{m}\rangle & = \sqrt{n} \vert n-1,\tilde{m}\rangle,\nonumber\\ 
a^\dagger\vert n,\tilde{m}\rangle &= \sqrt{n+1} \vert n+1,\tilde{m}\rangle,   \end{split}
\end{equation}
\begin{equation}
\begin{split}
\tilde{a}\vert n,\tilde{m}\rangle &= 
\sqrt{m} \vert n,\tilde{m}-1\rangle,\nonumber\\ 
\tilde{a}^\dagger\vert n,\tilde{m}\rangle &= \sqrt{m+1} \vert n,\tilde{m}+1\rangle.    
\end{split}
\end{equation}
The operators $a$ and $a^\dagger$  commute  with  $\tilde{a}$  and $\tilde{a}^\dagger$.  It is clear from the above that $a$ acts on the vector space $\cal{H}$ and $\tilde{a}$ acts on the vector space $\cal{H^*}$. From the expression for $\vert I\rangle$ in terms of the number states
\begin{equation}
\vert I\rangle = \sum_n \vert n,\tilde{n}\rangle,
\end{equation}
it follows that
\begin{equation}
a\vert I\rangle=\tilde{a}^\dagger \vert I\rangle,\; a^\dagger\vert I\rangle = 
\tilde{a}\vert I\rangle,
\end{equation}
and hence for any operator $A$ (written in terms of $a$ $a^\dagger$ and their complex conjugates), one has
\begin{equation}
A\vert I\rangle = \tilde{A}^\dagger \vert I\rangle,
\end{equation}
where $\tilde{A}$ is obtained from $A$ by making the  replacements 
tilde  conjugation  rules   $a\to   \tilde{a},   a^\dagger   \to 
\tilde{a}^\dagger,  \alpha\to  \alpha^*$. 
An immediate consequence of this is that  the state $\vert \rho^\alpha\rangle$
which remains unchanged under the replacements $a\to   \tilde{a}$,  
$a^\dagger  \to \tilde{a}^\dagger,$  and c number $\to$ complex conjugates C by applying the  the hermiticity property of $\rho$
i.e. $\rho^\dagger=\rho$. 
The tildian property reflects the hermiticity property of the density operator.

The non-equilibrium thermofield dynamics is developed and analysed in $\alpha=1$ representation. Hence, from now on, we work in, $\alpha=1$ representation \cite{shanta1996eigenstates,shanta1996operator,chaturvedi1999quantum}. In this representation, for any hermitian operator $A$, one has 
\begin{equation}
\langle A\rangle = \langle I\vert A\vert \rho\rangle = \langle A\vert\rho\rangle= Tr(A\rho).
\end{equation} 
The Von Neumann equation gives the evolution of a conservative system in terms of $\rho$, defined as:
\begin{eqnarray}
 \frac{\partial}{\partial t}\rho(t)=-\frac{i}{\hbar}[H,\rho].
\end{eqnarray}
By applying $\vert I\rangle$ from the right, one gets
\begin{equation*}
    i \frac{\partial}{\partial t} \rho \ket{I} = H \rho \ket{I} - \rho H \ket{I}.
\end{equation*}
In order to obtain the equation for $\ket{\rho} \equiv \hat{\rho} \ket{I}$, $\rho$ present in the second term of the R.H.S. of the above equation must move next to $\ket{I}$. This can be easily explained by the relations $a\vert I\rangle=\tilde{a}^\dagger \vert I\rangle,\; a^\dagger\vert I\rangle = \tilde{a}\vert I\rangle$ and the Hermiticity of $H$, which gives rise to the following equation:
\begin{eqnarray}
\frac{\partial}{\partial t}\vert \rho(t)\rangle = -i\hat{H}\vert \rho\rangle,
\end{eqnarray}
where
\begin{eqnarray}
 \hat{H}= (H-\tilde{H}),
\end{eqnarray}
and the Hamiltonian $H$ and $\tilde{H}$ belongs to the Hilbert space $\mathcal{H}$ and $\mathcal{H^*}$ respectively. With the application of TFD, one can derive a Schrödinger-like equation for any state $\vert\rho\rangle$ with an arbitrary value of $\alpha$.   
For dissipative systems, the evolution equation is given by the master equation, 
\begin{eqnarray}
 \frac{\partial}{\partial t}\rho(t)=-\frac{i}{\hbar}(H\rho-\rho H)+L\rho,
\end{eqnarray}
where $L$ is the Liouville term. The non-equilibrium thermofield dynamics is developed and analyzed in $\alpha=1$ representation. Hence, from now on, we work in $\alpha=1$ representation \cite{chaitanya2011algebraic, shanta1996eigenstates,shanta1996operator,chaturvedi1999quantum}.

By applying $\vert I\rangle$ to Eq. (8) from the right, one goes over to TFD, and the master equation is given by Eq. (9). 
With $-i\hat{H}$ in an Extended Hilbert space, the issue of solving a master equation is reduced to solving a Schrödinger-like equation.

In our paper, we convert the master equation associated with $su(1,1)$ symmetry into the Schrödinger-like equation using the TFD technique, and then apply the Hartree-Fock approximation to linearize the Hamiltonians. We analyze the entanglement and quantum mutual information for two different scenarios of a nonlinear master equation, followed by the disentanglement formula, single-mode and two-mode squeezed state.

\section{Nonlinear Quantum System}
We consider a model consisting of two single-modes separated by a distance $g$ and exposed to an external field $E(t)$, created by an external source. The two modes $a$ and $b$ are generated from a common source and thus share the same frequency. The Hamiltonian of the two single-mode system interacting with an external field is given by
\begin{equation}
    H = H_0 + H_1 = \epsilon_0 a^\dagger a + \epsilon_1 b^\dagger b +g_1( a^{\dagger} b E_1 + b^{\dagger} a E_1^*),
\end{equation}

\begin{figure}[ht]
    \centering
    \includegraphics[width=0.45\textwidth]{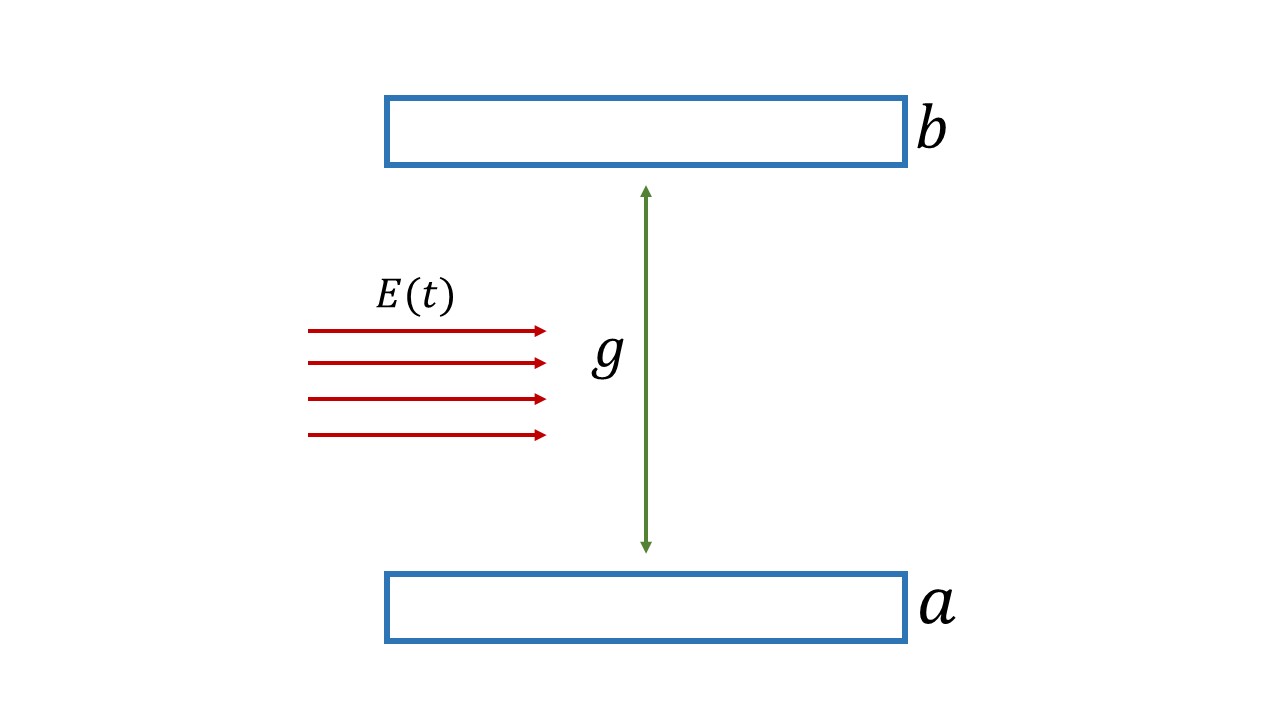}
    \caption{Schematic diagram of a coupled two single-mode system exposed to an external field $E(t)$, and $g$ is the coupling distance between them.}
    \label{fig:gamma}
\end{figure}

where $H_0 = \epsilon_0 a^\dagger a + \epsilon_1 b^\dagger b$, $H_1 = g_1( a^{\dagger} b E_1 + b^{\dagger} a E_1^*)$ and
the Hamiltonian without the external field $E(t)$ is the quantum system described in the reference \cite{rai2010quantum}. The first two terms in the Hamiltonian containing $\epsilon_0$ and $\epsilon_1$ are the energies of the modes $a$ and $b$ respectively, and the last two terms arise due to the coupling of two single modes, with $ g$ referring to the distance between these two modes, which are real. Also, $a,b$ and $ a^{\dagger},b^\dagger$ are the bosonic annihilation and creation operators, satisfying the commutation relation $[a_{i}, a_{j}^{\dagger}] = \delta_{ij}$ and the external field $E_1(t)$, is arised from an external source, where $E_1(t) = \Bar{E}_1(t) \exp(- iw_1t)$. $\Bar{E}_1$ specifies the amplitude of the external field.\\
The master of the corresponding system is given by 
\begin{equation}
    \frac{d}{dt} \rho = \frac{1}{i \hbar} [H, \rho] + \frac{d}{dt} \rho_{irr},
    \end{equation}
where $\frac{d}{dt} \rho_{irr}$ refers to the dissipation in the entire system \cite{chaturvedi1991solution}, given by
\begin{equation}
\frac{d}{dt} \rho_{irr}=\frac{\gamma_1}{2} [a^{\dagger} b \rho, b^{\dagger} a] + \frac{\gamma_1}{2} [a^{\dagger} b, \rho b^{\dagger} a].
\end{equation}
Applying $\ket{I}$ on both sides of Eq. (13), the TFD notation goes over to the following notation
\begin{equation}
    \frac{d}{dt} \ket{\rho} = \left( \frac{1}{i \hbar} [H, \rho] + \frac{\gamma_1}{2} [a^{\dagger} b \rho, b^{\dagger} a] + \frac{\gamma_1}{2} [a^{\dagger} b, \rho b^{\dagger} a] \right) \ket{I},
\end{equation}
where $\ket{\rho}$ is a state vector in an extended Hilbert space $\mathcal{H} \otimes \mathcal{H}^*$.
It is clear from the above equation that Eq. (13) is now reduced to Eq. (9), which is of the form
\begin{equation}
{ \frac{\partial}{\partial t} \ket{\rho} = -i \hat{H_1} \ket{\rho}},
\end{equation}
where
\begin{equation}
\begin{split}
-i\hat{H_1} = &\frac{1}{i \hbar} \Bigl[\frac{\epsilon_0}{2}(a^{\dagger} a - \tilde{a} \tilde{a}^{\dagger}) + \epsilon_1 (b^{\dagger} b - \tilde{b} \tilde{b}^{\dagger})
+ g_1 \biggl\{ (a^{\dagger} b - \tilde{a} \tilde{b}^{\dagger}) E_1 + (b^{\dagger} a - \tilde{b} \tilde{a}^{\dagger}) E_1^* \biggr\} \Bigr] \\
& + \frac{\gamma_1}{2} \biggl( 2a^{\dagger} b \tilde{b} \tilde{a}^{\dagger} + 2b^{\dagger} a \tilde{a} \tilde{b}^{\dagger} - b^{\dagger} a a^{\dagger} b - a^{\dagger} b b^{\dagger} a - \tilde{b} \tilde{a}^{\dagger} \tilde{a} \tilde{b}^{\dagger} - \tilde{a} \tilde{b}^{\dagger} \tilde{b} \tilde{a}^{\dagger} \biggr). 
\end{split}
\end{equation}

Here, the annihilation and creation operators $a, b, {a}^\dagger$ and ${b}^\dagger$ act on the Hilbert space $\mathcal{H}$; on the other hand, the operators $\tilde{a}, \tilde{b}, \tilde{a}^{\dagger}$ and $\tilde{b}^{\dagger}$ are acting on the Hilbert space $\mathcal{H}^*$. Now, it should be clear to the reader that the master equation in Eq. (13) is reduced to the Schrödinger-like Eq. (16) in the extended Hilbert space.  

The presence of the dissipation like, $(\tilde{b} \tilde{a}^{\dagger} \tilde{a} \tilde{b}^{\dagger} - \tilde{a} \tilde{b}^{\dagger} \tilde{b} \tilde{a}^{\dagger})$ terms in Eq. (17) gives rise to nonlinearity \cite{ottinger2010nonlinear,ottinger2011geometry, shanta1996eigenstates}. Applying the Hartree-Fock approximation \cite{chaturvedi1990use,shiv2014entanglement} to each term of Eq. (17) gives (details are given in Appendix B) 
\begin{equation}
\begin{split}
& \langle a^{\dagger} b \rangle = \langle b^{\dagger} a \rangle = \langle \tilde{a}^{\dagger} \tilde{b} \rangle = \langle \tilde{b}^{\dagger} \tilde{a} \rangle = \Delta_1 ;\;\;
\langle ab \rangle = \langle a^{\dagger} b^{\dagger} \rangle = \langle \tilde{a} \tilde{b} \rangle = \langle \tilde{a}^{\dagger} \tilde{b}^{\dagger} \rangle = \Delta . 
\end{split}   
\end{equation}
 The Hamiltonian in Eq. (17) can be written as tilden and non-tilden by considering $E_1 = E_1^*$, which is given by
\begin{equation}
-i \hat{H_1} = -i (H_1 - \tilde{H}_1),  
\end{equation}
where
\begin{equation}
\begin{split}
 -iH_1 = \Bigl( \frac{\epsilon_0}{2i\hbar} - \frac{\gamma_1}{2} \Bigr) a^{\dagger} a + \Bigl( \frac{\epsilon_1}{i\hbar} - \frac{\gamma_1}{2} \Bigr) b^{\dagger} b + \Bigl( \frac{g_1 E_1}{i \hbar} + \frac{\gamma_1 \Delta_1}{2} \Bigr) \Bigl( a^{\dagger} b + b^{\dagger} a \Bigr) - \frac{\gamma_1 \Delta}{2} \Bigl( ab + a^{\dagger} b^{\dagger} \Bigr),   
\end{split}
\end{equation}
and
\begin{equation}
    \begin{split}
i\tilde{H}_1 = - \Bigl( \frac{\epsilon_0}{2i\hbar} - \frac{\gamma_1}{2} \Bigr) \tilde{a}^{\dagger} \tilde{a} - \Bigl( \frac{\epsilon_1}{i\hbar} - \frac{\gamma_1}{2} \Bigr) \tilde{b}^{\dagger} \tilde{b} - \Bigl( \frac{g_1 E_1}{i \hbar} + \frac{\gamma_1 \Delta_1}{2} \Bigr) \Bigl( \tilde{a}^{\dagger} \tilde{b} + \tilde{b}^{\dagger} \tilde{a} \Bigr) - \frac{\gamma_1 \Delta}{2} \Bigl( \tilde{a} \tilde{b} + \tilde{a}^{\dagger} \tilde{b}^{\dagger} \Bigr).
    \end{split}
\end{equation}
In our model, the modes $a$ and $b$ share the same frequency variables, which makes $\Delta_1$ to be time-independent. Considering $\Bigl( \frac{\epsilon_0}{2i\hbar} - \frac{\gamma_1}{2} \Bigr) = r$, $\Bigl( \frac{\epsilon_1}{i\hbar} - \frac{\gamma_1}{2} \Bigr) = s$ and $\Bigl( \frac{g_1 E_1}{i \hbar} + \frac{\gamma_1 \Delta_1 }{2} \Bigr) = \eta =\eta^*$, the Hamiltonian of Eq. (20) and Eq. (21) can be written as
\begin{equation}
-iH_1 = rs (a^{\dagger} a + b^{\dagger} b) + \eta \sqrt{rs} (a^{\dagger} b + b^{\dagger} a) - \frac{\gamma_1 \Delta}{2} \sqrt{rs} (ab + a^{\dagger} b^{\dagger}),
\end{equation}
and
\begin{equation}
i\tilde{H}_1 =-rs (\tilde{a}^{\dagger} \tilde{a} + \tilde{b}^{\dagger} \tilde{b}) - \eta^* \sqrt{rs} (\tilde{a}^{\dagger} \tilde{b} + \tilde{b}^{\dagger} \tilde{a}) - \frac{\gamma_1 \Delta}{2} \sqrt{rs} (\tilde{a} \tilde{b} + \tilde{a}^{\dagger} \tilde{b}^{\dagger}).
\end{equation}
In the following subsections, we compute the squeezing parameter, followed by the disentanglement theorem, leading to the derivation of entanglement and quantum mutual information for two distinct scenarios, such as $\eta =0$ and $\eta \ne 0$.

\subsection{Case-1: If \texorpdfstring{$\eta = 0$}{eta=0}}
For subsequent computation, we are taking into account the non-tilden Hamiltonian, calculated in Eq. (22). The non-tilden Hamiltonian consists of both the external field and the Hartree-Fock field. When the Hartree-Fock field is comparable to the external field, i.e, $ \frac{g_1 E_1}{i \hbar} = - \frac{\gamma_1 \Delta_1}{2}$, it leads to $\eta =0 $ and Eq. (22) being restated as
\begin{equation}
-iH_1 = rs (a^{\dagger} a + b^{\dagger} b) - \frac{\gamma_1 \Delta}{2} \sqrt{rs} (ab + a^{\dagger} b^{\dagger}),    
\end{equation}
which is diagonalized by the following transformations among the mode operators \cite{shiv2014entanglement,chalker2013quantum,birol2007bogoliubov,emary2001bogoliubov}
\begin{equation}
\begin{split}
& A = \mu a + \nu^* b^{\dagger}, A^{\dagger}  = \mu^* a^{\dagger} + \nu b,\\
& B = \mu b + \nu^* a^{\dagger}, B^{\dagger}  = \mu^* b^{\dagger} + \nu a,
\end{split}
\end{equation}
with $\mu= \mu^* = \cosh{\textbf{r}}$ and $\nu^* = \sinh{\textbf{r}} e^{i \phi}, \nu = \sinh{\textbf{r}} e^{-i \phi}$. This transformation is nothing but identical to the evolution of the operators, defined as \cite{rai2010quantum}
\begin{equation}\begin{split}
& A = S^\dagger(\xi) a S(\xi) = \cosh(\textbf{r}) a + e^{i \phi} \sinh (\textbf{r}) b^{\dagger}, A^{\dagger}  = \cosh(\textbf{r}) a^{\dagger} + e^{-i \phi} \sinh (\textbf{r}) b,\\
& B = S^\dagger(\xi) b S(\xi) = \cosh(\textbf{r}) b + e^{i \phi} \sinh (\textbf{r}) a^{\dagger}, B^{\dagger}  = \cosh(\textbf{r}) b^{\dagger} + e^{-i \phi} \sinh (\textbf{r}) a.
\end{split}
\end{equation}
This transformation is similar to the Bogoliubov transformation \cite{suzuki2011lecture} given in Eq. (25) and 
\begin{equation}
S(\xi) = \exp (\xi a^{\dagger} b^{\dagger} - \xi^* ab), \xi = \textbf{r} e^{i \phi},
\end{equation}
where $\textbf{r}$ is the squeezing parameter. In order to study the entanglement properties of the two-mode squeezed state, we have defined the Hamiltonian in terms of $su(1,1)$ operators, which satisfy the $su(1,1)$ algebra, or the Lie algebra, is given by 
\begin{equation}
\begin{split}
-iH_1 & = rs K_3 - \frac{\gamma_1 \Delta}{2} \sqrt{rs} (K_+ + K_-)- 2 rs,
\end{split}
\end{equation}
which can be diagonalized to the new Hamiltonian $-iH_{1f}$ after the unitary transformation of $-iH_1$ defined as:
\begin{equation}
-iH_{1f} = S^\dagger (\xi) (-iH_1) S(\xi),
\end{equation}
with $S(\xi) = \exp (\xi K_+ - \xi^* K_-), \xi = \textbf{r} e^{i \phi}$.
Similar method is applicable for the tilden part of the Hamiltonian.\\
The solution to the Schrödinger equation in Eq. (16) is given by
\begin{equation*}
 \ket{\rho(t)} = [e^{-i \int H_1 dt} \otimes e^{-i \int \tilde{H}_1 dt}] \ket{\rho (0)},
\end{equation*}
where, $\hat{H}_1 = H_1 + \tilde{H}_1$ and $\ket{\rho(0)}$ is an initial state in $\mathcal{H} \otimes \mathcal{H}^*$. From the above equation, it is clear that the two Hamiltonians $H_1$ and $\tilde{H}_1$ are independent in the sense that $H_1$ will act on a non-tilden system and $\tilde{H}_1$ will act on a tilden system. Using the disentanglement theorem \cite{wodkiewicz1985coherent}, the above equation can be simplified as
\begin{equation}
\begin{split}
\ket{\rho(t)} & = e^{-i \int[rs K_3 - \frac{\gamma_1 \Delta}{2} \sqrt{rs} (K_+ + K_-)- 2rs] dt} \otimes e^{-i \int[ rs \tilde{K}_3 - \frac{\gamma_1 \Delta}{2} \sqrt{rs} (\tilde{K}_+ + \tilde{K}_-)- 2rs]dt} \ket{\rho(0)}\\
& = \exp [\xi_3 K_3 + \xi_+ K_+ + \xi_- K_-] \otimes \exp [\tilde{\xi}_3 \tilde{K}_3 + \tilde{\xi}_+ \tilde{K}_+ + \tilde{\xi}_- \tilde{K}_-]. \exp[-4rst] \ket{\rho(0)}\\
& = \exp[\Gamma_+ K_+] \exp[\ln{\Gamma_3 K_3}] \exp[\Gamma_- K_-] \otimes \exp[\tilde{\Gamma}_+ \tilde{K}_+] \exp[\ln{\tilde{\Gamma}_3 \tilde{K}_3}] \exp[\tilde{\Gamma}_- \tilde{K}_-] \times \\
& \exp[{-4rst}] \ket{\rho(0)},
\end{split}
\end{equation}
where
\begin{equation}
\begin{split}
& \Gamma_{\pm} = \frac{2 \xi_{\pm} \sinh{\phi}}{2 \phi \cosh{\phi} - \xi_{3} \sinh{\phi}},\Gamma_{3} = \frac{1} {\bigl(\cosh{\phi} - \frac{\xi_{3}}{2 \phi} \sinh{\phi}\bigr)^2}, \phi^2 = \frac{\xi^2_{3}}{4} - \xi_{+} \xi_{-}. 
\end{split}
\end{equation}
Considering a general, arbitrary initial condition for $\rho (0)= \sum_{m,n,m',n'}^\infty \rho_{m,n,m',n'} (0,0) \ket{m,m'} \bra{n,n'}$, where $\ket{m,m'}$ and $\ket{n,n'}$ indicates the number states. In thermofield dynamic notation, $\ket{\rho(0)}$ takes the form, $\ket{\rho (0)}= \sum_{m,n,m',n'}^\infty \rho_{m,n,m',n'} (0,0) \ket{m,n,m',n'}$. Now, the successive action of the operators on the initial state in Eq. (24) can be expressed as
\begin{equation}
\begin{split}
\ket{\rho(t)} = & \sum_{u=0}^{min (m,n)} \sum_{v=0}^{\infty} {\Biggl[ \left( \begin{array}{c} m+u-v \\ u \end{array} \right) \left( \begin{array}{c} n+u-v \\ u \end{array} \right) \left( \begin{array}{c} m \\ v \end{array} \right) \left( \begin{array}{c} n \\ v \end{array} \right) \Biggr]}^{\frac{1}{2}}\\
& \times \sum_{u'=0}^{min (m',n')} \sum_{v'=0}^{\infty} {\Biggl[ \left( \begin{array}{c} m'+u'-v' \\ u' \end{array} \right) \left( \begin{array}{c} n'+u'-v' \\ u' \end{array} \right) \left( \begin{array}{c} m' \\ v' \end{array} \right) \left( \begin{array}{c} n'\\ v' \end{array} \right) \Biggr]}^{\frac{1}{2}}\\
& \times {\bigl[ \Gamma_+ \bigr]}^u {\bigl[ \Gamma_3 \bigr]}^{(m+n-2v+1)/2} {\bigl[ \Gamma_- \bigr]}^v \times {\bigl[ \tilde{\Gamma}_+ \bigr]}^{u'} {\bigl[ \tilde{\Gamma}_3 \bigr]}^{(m'+n'-2v'+1)/2} {\bigl[ \tilde{\Gamma}_- \bigr]}^{v'} \exp[{-4rst}]\\
& \times \ket{\rho_{m+u-v,n+u-v,m'+u'-v',n'+u'-v'} (0)},
\end{split}
\end{equation}

\begin{equation}
\begin{split}
\Rightarrow \ket{\psi(t)} = & \sum_{u=0}^{min (m,n)} \sum_{v=0}^{\infty} {\Biggl[ \left( \begin{array}{c} m+u-v \\ u \end{array} \right) \left( \begin{array}{c} n+u-v \\ u \end{array} \right) \left( \begin{array}{c} m \\ v \end{array} \right) \left( \begin{array}{c} n \\ v \end{array} \right) \Biggr]}^{\frac{1}{2}}\\
& \times {\bigl[ \Gamma_{+} \bigr]}^u {\bigl[ \Gamma_{3} \bigr]}^{(m+n-2v+1)/2} {\bigl[ \Gamma_{-} \bigr]}^v \exp[{-2rst}] \times \ket{\rho_{m+u-v,n+u-v} (0)}.
\end{split}
\end{equation}
We can construct a density matrix, defined as
\begin{equation}
\rho(t) = \ket{\psi} \bra{\psi}.
\end{equation}
Applying the TFD technique to the above equation by multiplying $\ket{I}$, we have
\begin{equation}
\rho \ket{I} = \ket{\psi} \bra{\psi} \ket{I} \Rightarrow \ket{\rho(t)} = \ket{\psi, \tilde{\psi}}.
\end{equation}
Now Eq. (35) is identical to Eq. (32), which results in the decoupling of the system $ab$ and $\tilde{a}\tilde{b}$.
Finally, the wave function is of the form
\begin{equation}
\begin{split}
\ket{\psi(t)} = &  \exp[\Gamma_+ K_+] \exp[\ln{\Gamma_3 K_3}] \exp[\Gamma_- K_-]  \times \exp[{-2rst}] \ket{\rho(0)},
\end{split}
\end{equation}
and $\xi_3 = \int rs dt, \xi_+ = \xi_-= - \int \frac{\gamma_1 \Delta}{2} \sqrt{rs} dt$.
Here, we can clearly see that $\Gamma$'s are the function of $\Delta$, which can be calculated by using Eq. (31) and from the expression of $\Delta$, as (see Appendix D for more details)
\begin{equation}
\Gamma_{\pm} = -\frac{\gamma_1^2 \mathcal{W}^2}{4} \Delta t (1+\frac{ \mathcal{W}t}{2}) e^{\frac{-i \mathcal{W}t}{2}},
\end{equation}
and
\begin{equation}
\textbf{r} = \frac{\gamma_1^2 \mathcal{W}^2}{4} \Delta t (1+\frac{ \mathcal{W}t}{2}), 
\end{equation} 
where $\mathcal{W} = rs$.

\subsubsection{Entanglement Calculation}
In this section, we will analyse the entanglement evolution of two-mode squeezed states. For this, we have considered the two-mode squeezed state as
\begin{equation}
\ket{\xi} = \exp (\xi a^{\dagger} b^{\dagger} - \xi^* ab) \ket{0,0}, \xi = \textbf{r} e^{i \phi}.
\end{equation}
where $\ket{\xi} = S(\xi) \ket{0,0} = \exp (\xi a^{\dagger} b^{\dagger} - \xi^* ab) \ket{0,0}$. Then we will construct a covariance matrix $\sigma$ for the operator $S(\xi)$ as \cite{rai2010quantum}
\begin{equation}
\sigma=
\begin{pmatrix}
f & g & h & 0\\
g & f & 0 & -h\\
h & 0 & f & g\\
0 & -h & g & f
\end{pmatrix}
= 
\begin{pmatrix}
A & C\\
C^T & B
\end{pmatrix},
\end{equation}
where
\begin{equation}
\begin{split}
& f= \frac{1}{2} \cosh(2 \textbf{r}),\;
 g= -\frac{1}{2} \sinh(2 \textbf{r}) \sin(2 \phi),\;
h = \frac{1}{2} \sinh(2 \textbf{r}) \cos(2 \phi), 
\end{split}
\end{equation}
which leads to the calculation of logarithmic negativity $E(\sigma)$ for a two-mode Gaussian state given by \cite{vidal2002computable}
\begin{equation}
 E(\tilde{\sigma}) = max\{ 0,-\log (2\tilde{d_-})\},   
\end{equation}
which is a simple increasing function of the minimum symplectic eigenvalue $\tilde{d}_-$ of the transpose of the covariance matrix $\sigma$. Hence, it provides a fine method for evaluating entanglement. The eigenvalues of the covariance matrix are given by
\begin{equation}
\begin{split}
\tilde{d}_{\pm} & = \sqrt{\frac{\tilde{\Delta}(\sigma) \pm \sqrt{\tilde{\Delta}(\sigma)^2 - 4 Det(\sigma)}}{2}},
\end{split}
\end{equation}
where
\begin{equation}
 \tilde{\Delta}(\sigma) = Det(A) + Det(B) - 2 Det(C).
\end{equation}
\begin{figure}[ht]
    \centering
    \includegraphics[width=0.75\textwidth]{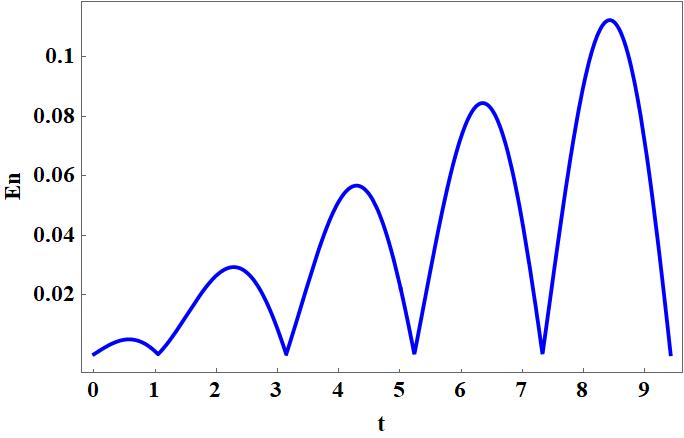}
    \caption{Time evolution of logarithmic negativity $E_n$ with respect to $t$. Here $\gamma_1$ = 0.1, $\mathcal{W}$ = 1.5 and $\Delta$ = 0.1. }
    \label{fig:eta1}
\end{figure}
From the expressions of Eqs. (38), (42), and (43), the eigenvalues are dependent on the squeezing parameter, and from our calculation, the squeezing parameter is a function of time. By plotting the graph of logarithmic negativity versus time in Fig. \ref{fig:eta1}, it can be clearly seen that, with an increase in time, the entanglement also increases continuously. At $t=0$, $E_n=0$, and the state, $\ket{\xi}$ is separable. With increasing t value, it oscillates between zero and non-zero values, indicating the state, $\ket{\xi}$, becomes entangled and disentangled with growing time, which is initially separable.

\subsubsection{Quantum Mutual Information Calculation}
The expression for quantum mutual information is given as \cite{olivares2012quantum,slepian2003noiseless}:
\begin{equation}
I_M (\rho_{AB}) = S_V(\rho_A) + S_V(\rho_B) - S_V(\rho_{AB}),
\end{equation}
where $I_M$ denotes the quantum mutual information and $S_V(\rho) = - Tr[\rho \ln \rho]$, known as the von Neumann entropy. The above expression can be easily rewritten in terms of symplectic eigenvalues as:
\begin{equation}
 I_M (\rho_{AB}) = f(\sqrt{Det A}) + f(\sqrt{Det B}) - f(d_+) - f(d_-),
\end{equation}
where $d_{\pm}$ are the symplectic eigenvalues of the matrix $\sigma$ and  $f(x) = (x+\frac{1}{2}) \ln (x+\frac{1}{2}) - (x-\frac{1}{2}) \ln (x-\frac{1}{2})$. 

\begin{figure}[ht]
    \centering
    \includegraphics[width=0.75\textwidth]{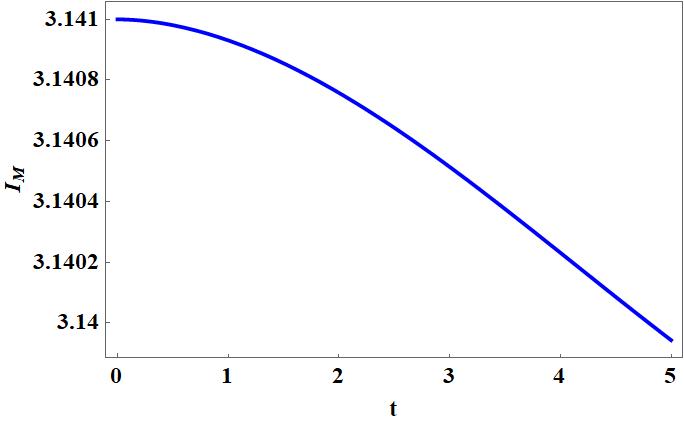}
    \caption{Time evolution of quantum mutual information $I_m$ with respect to $t$. Here, $\gamma_1$ = 0.1, $\mathcal{W}$ = 0.1 and $\Delta$ = 0.1. }
    \label{fig:eta2}
\end{figure}
The entanglement and mutual information calculation is done for an approximated value of the Hartree-Fock field $\Delta$. We calculate the  numerical value of $\Delta$ without approxmation upto order $t$ using the system parameters: $\gamma_1$ = 0.1, $\mathcal{W}= rs = 1.5 $ rad/s and $\Delta(0)$ = 0.1 (as mentioned in Fig. \ref{fig:eta1}), we obtain
\begin{equation*}
\Delta = 0.1 c + 0.0118184 - i t 0.0115786,
\end{equation*}
for temperature, $T = 10^{-11} K$ and states $m,n$ varies from 1 to 2. Here, we have calculated the $\Delta$ value in the order of $t$. The coefficients of higher-order t will give very small values and can be neglected. Here, for the numerical value of $\Delta$, we got the temperature $T= 10^{-11} K$, and for different frequencies, the temperature will change.

\subsection{Case-2: If \texorpdfstring{$\eta  \ne 0$}{eta ne 0}}
From our calculation, it is seen that the non-tilden Hamiltonian in Eq. (22) consists of both the external field and the Hartree-Fock field. In this case, we consider that both of the fields are not comparable, which results in $\eta \ne 0$. So, the Hamiltonian in Eq. (22) is of the form
\begin{equation}
-iH_1 = rs (a^{\dagger} a + b^{\dagger} b) + \eta \sqrt{rs} (a^{\dagger} b + b^{\dagger} a) - \frac{\gamma_1 \Delta(t)}{2} \sqrt{rs} (ab + a^{\dagger} b^{\dagger}),   
\end{equation}
which is modified into a new Hamiltonian,
\begin{equation}
\begin{split}
    -iH'_1 = & \sqrt{rs} (\sqrt{rs} + \eta) (A^{\dagger} A +  A A^{\dagger}) - \frac{\gamma_1 \Delta}{2} \sqrt{rs} (A^2 + A^{\dagger 2})\\
   + &  \sqrt{rs} (\sqrt{rs} - \eta) (B^{\dagger} B +  B B^{\dagger}) + \frac{\gamma_1 \Delta}{2} \sqrt{rs} (B^2 + B^{\dagger 2}) - 2rs,
\end{split}
\end{equation}
through the following transformation:
\begin{equation}
\begin{split}
& a = A + B, a^\dagger = A^\dagger + B^\dagger,\\
& b = A - B, b^\dagger = A^\dagger - B^\dagger.
\end{split}
\end{equation}
The evolution of operators is defined as \cite{shiv2014entanglement},
\begin{equation}
\begin{split}
& X = \mu A + \nu^* A^{\dagger}, X^{\dagger}  = \mu^* A^{\dagger} + \nu A,\\
& Y = \mu B + \nu^* B^{\dagger}, Y^{\dagger}  = \mu^* B^{\dagger} + \nu B,
\end{split}
\end{equation}
where, $\mu= \mu^* = \cosh{\mathbf{r}}$ and $\nu^* = \sinh{\mathbf{r}} e^{i \phi}, \nu = \sinh{\mathbf{r}} e^{-i \phi}$. This transformation is the Bogoliubov transformation \cite{shiv2014entanglement,chalker2013quantum,birol2007bogoliubov,emary2001bogoliubov}, which diagonalizes the Hamiltonian. Defining the Hamiltonian in terms of $su(1,1)$ operators, we have
\begin{equation}
\begin{split}
    -i\mathscr{H}_1 = & \sqrt{rs} (\sqrt{rs} + \eta) \mathcal{N}_{a3} - \frac{\gamma_1 \Delta}{2} \sqrt{rs} (K_{a-} + K_{a+})\\
    + & \sqrt{rs} (\sqrt{rs} - \eta) \mathcal{N}'_{b3} + \frac{\gamma_1 \Delta}{2} \sqrt{rs} (K_{b-} + K_{b+}) - 2rs,
\end{split}
\end{equation}
by taking 
\begin{equation}
\begin{split}
& \mathcal{N}_{a3} = A^{\dagger}A + A A^{\dagger}, K_{a+} = A^{\dagger} A^{\dagger},K_{a-} = AA,\\
& \mathcal{N'}_{b3} = B^{\dagger}B + B B^{\dagger}, K'_{b+} = B^{\dagger} B^{\dagger},K'_{b-} = BB, 
\end{split}
\end{equation}
which satisfies the $su(1,1)$ algebra or Lie algebra:
\begin{equation}
[\mathcal{N}, K_{+}] = K_{+}, [\mathcal{N}, K_{-}] = -K_{-} \hspace{1mm}\text{and} \hspace{1mm} [K_{-}, K_{+}] = 2\mathcal{N}.
\end{equation}
And 
\begin{equation}
S(\xi) = \exp (\xi A^{\dagger2}- \xi^* A^2), \xi = \textbf{r} e^{i \phi},
\end{equation}
where $\textbf{r}$ is the squeezing parameter. Similar method is applicable to diagonalize the tilden part of the Hamiltonian. Now, the Schrödinger equation can be rewritten into a simplified form as
\begin{equation}
\begin{split}
   &  \frac{\partial}{\partial t} \ket{\rho(t)} = -i \hat{H}_1 \ket{\rho},\\ \Rightarrow  & \ket{\rho(t)} = [e^{-i \int\mathscr{H}_1 dt} \otimes e^{-i \int \tilde{\mathscr{H}_1} dt}] \ket{\rho (0)},\\ 
  = & \bigl[ e^{\int [\sqrt{rs} (\sqrt{rs} + \eta)\mathcal{N}_{a3} - \frac{\gamma_1 \Delta}{2} \sqrt{rs} (K_{a-} + K_{a+})] dt} \otimes e^{\int [\sqrt{rs} (\sqrt{rs} - \eta)\mathcal{N}_{b3} + \frac{\gamma_1 \Delta}{2} \sqrt{rs} (K_{b-} + K_{b+})] dt} . e^{-2rst} \\
& \otimes e^{ \int [\sqrt{rs} (\sqrt{rs} + \eta')\mathcal{\tilde{N}}_{a3} - \frac{\gamma_1 \Delta}{2} \sqrt{rs} (\tilde{K}_{a-} + \tilde{K}_{a+})] dt} \otimes e^{\int [\sqrt{rs} (\sqrt{rs} - \eta')\mathcal{\tilde{N}}_{b3} + \frac{\gamma_1 \Delta}{2} \sqrt{rs} (\tilde{K}_{b-} + \tilde{K}_{b+})] dt} . e^{-2rst} \bigr] \ket{\rho(0)},\\
= & \exp[\xi_{a3} \mathcal{N}_{a3} + \xi_{a+} K_{a+} + \xi_{a-} K_{a-}] \otimes \exp[\xi_{b3} \mathcal{N}_{b3} + \xi_{b+} K_{b+} + \xi_{b-} K_{b-}]\\
& \otimes \exp[\xi'_{a3} \mathcal{\tilde{N}}_{a3} + \xi'_{a+} \tilde{K}_{a+} + \xi'_{a-} \tilde{K}_{a-}] \otimes \exp[\xi'_{b3} \mathcal{\tilde{N}}_{b3} + \xi'_{b+} \tilde{K}_{b+} + \xi'_{b-} \tilde{K}_{b-}] . e^{-2rst} . e^{-2rst} \ket{\rho(0)},
\end{split}
\end{equation}
where $\ket{\rho(0)}$ is an initial state in $\mathcal{H} \otimes \mathcal{H}^*$ and
\begin{equation}
\begin{split}
    & \xi_{a3} = \int dt \sqrt{rs} (\sqrt{rs} + \eta), \xi_{a+} = \xi_{a-} = - \int dt \frac{\gamma_1 \Delta}{2} \sqrt{rs},\\
    & \xi_{b3} = \int dt \sqrt{rs} (\sqrt{rs} - \eta), \xi_{b+} = \xi_{b-} =  \int dt \frac{\gamma_1 \Delta}{2} \sqrt{rs,}\\
    & \xi'_{a3} = \int dt \sqrt{rs} (\sqrt{rs} + \eta'), \xi'_{a+} = \xi'_{a-} = - \int dt \frac{\gamma_1 \Delta}{2} \sqrt{rs},\\
    & \xi'_{b3} = \int dt \sqrt{rs} (\sqrt{rs} - \eta'), \xi_{b+} = \xi_{b-} = \int dt \frac{\gamma_1 \Delta}{2} \sqrt{rs}.
\end{split}
\end{equation}
Using the disentanglement theorem \cite{wodkiewicz1985coherent} in Eq. (55), we have
\begin{equation}
\begin{split}
\ket{\rho(t)} = & \bigl[ \exp[\Gamma_{a+} K_{a+}] \exp[ln(\Gamma_{a3} \mathcal{N}_{a3})] \exp[\Gamma_{a-} K_{a-}]
\otimes \exp[\Gamma_{b+} K_{b+}] \exp[ln(\Gamma_{b3} \mathcal{N}_{b3})] \exp[\Gamma_{b-} K_{b-}]\\
\otimes & \exp[\Gamma'_{a+} \tilde{K}_{a+}] \exp[ln(\Gamma'_{a3} \mathcal{\tilde{N}}_{a3})] \exp[\Gamma'_{a-} \tilde{K}_{a-}] 
\otimes \exp[\Gamma'_{b+} \tilde{K}_{b+}] \exp[ln(\Gamma'_{b3} \mathcal{\tilde{N}}_{b3})] \exp[\Gamma'_{b-} \tilde{K}_{b-}]\\
 & e^{-2rst} e^{-2rst} \bigr] \ket{\rho(0)},
\end{split}
\end{equation}
where
\begin{equation}
\begin{split}
\Gamma_{i \pm} = \frac{2 \xi_{i\pm} \sinh{\phi_i}}{2 \phi_i \cosh{\phi_i} - \xi_{i3} \sinh{\phi_i}} = \Gamma'_{i\pm}, \Gamma_{i3} = \frac{1} {\bigl(\cosh{\phi_i} - \frac{\xi_{i3}}{2 \phi_i} \sinh{\phi_i}\bigr)^2} = \Gamma'_{i3},\phi^2_i = \frac{\xi^2_{i3}}{4} - \xi_{i+} \xi_{i-}, 
\end{split}
\end{equation}
and $i$ stands for $a$ and $b$. In thermofield dynamic notation, the initial state $\ket{\rho(0)}$ can be written as, $\ket{\rho (0)}= \sum_{m,n,m',n'}^\infty \rho_{m,n,m',n'} (0,0) \ket{m,n,m',n'}$ and with the successive action of the operators on the initial state in the above equation can be written as
\begin{equation}
\begin{split}
\ket{\rho(t)} = & \sum_{u=0}^{min (m,n)} \sum_{v=0}^{\infty} {\Biggl[ \left( \begin{array}{c} m+u-v \\ u \end{array} \right) \left( \begin{array}{c} n+u-v \\ u \end{array} \right) \left( \begin{array}{c} m \\ v \end{array} \right) \left( \begin{array}{c} n \\ v \end{array} \right) \Biggr]}^{\frac{1}{2}}\\
& \times \sum_{u'=0}^{min (m',n')} \sum_{v'=0}^{\infty} {\Biggl[ \left( \begin{array}{c} m'+u'-v' \\ u' \end{array} \right) \left( \begin{array}{c} n'+u'-v' \\ u' \end{array} \right) \left( \begin{array}{c} m' \\ v' \end{array} \right) \left( \begin{array}{c} n'\\ v' \end{array} \right) \Biggr]}^{\frac{1}{2}}\\
& \times {\bigl[ \Gamma_+ \bigr]}^u {\bigl[ \Gamma_3 \bigr]}^{(m+n-2v+1)/2} {\bigl[ \Gamma_- \bigr]}^v \times {\bigl[ \tilde{\Gamma}_+ \bigr]}^{u'} {\bigl[ \tilde{\Gamma}_3 \bigr]}^{(m'+n'-2v'+1)/2} {\bigl[ \tilde{\Gamma}_- \bigr]}^{v'} \exp[{-4rst}]\\
& \times \ket{\rho_{m+u-v,n+u-v,m'+u'-v',n'+u'-v'} (0)},
\end{split}
\end{equation}

\begin{equation}
\begin{split}
\Rightarrow \ket{\psi(t)} = & \sum_{u=0}^{min (m,n)} \sum_{v=0}^{\infty} {\Biggl[ \left( \begin{array}{c} m+u-v \\ u \end{array} \right) \left( \begin{array}{c} n+u-v \\ u \end{array} \right) \left( \begin{array}{c} m \\ v \end{array} \right) \left( \begin{array}{c} n \\ v \end{array} \right) \Biggr]}^{\frac{1}{2}}\\
& \times {\bigl[ \Gamma_{+} \bigr]}^u {\bigl[ \Gamma_{3} \bigr]}^{(m+n-2v+1)/2} {\bigl[ \Gamma_{-} \bigr]}^v \exp[{-2rst}] \times \ket{\rho_{m+u-v,n+u-v} (0)}.
\end{split}
\end{equation}
We can construct a density matrix, defined as
\begin{equation}
\rho(t) = \ket{\psi} \bra{\psi}.
\end{equation}
Since in the TFD technique $\ket{\rho(t)}$ can be written as
\begin{equation}
\rho \ket{I} = \ket{\psi} \bra{\psi} \ket{I} \Rightarrow \ket{\rho(t)} = \ket{\psi, \tilde{\psi}},
\end{equation}
then
\begin{equation}
\begin{split}
\ket{\psi(t)} = & \bigl[ \exp[\Gamma_{a+} K_{a+}] \exp[ln(\Gamma_{a3} \mathcal{N}_{a3})] \exp[\Gamma_{a-} K_{a-}]\\
\otimes & \exp[\Gamma_{b+} K_{b+}] \exp[ln(\Gamma_{b3} \mathcal{N}_{b3})] \exp[\Gamma_{b-} K_{b-}] e^{-2rst} \bigr] \ket{\rho(0)}.
\end{split}
\end{equation}
Here, $\Gamma$'s are the function of $\Delta$, which can be calculated by using Eq. (58) and from the expression of $\Delta$ (see Appendix F, for more details) as
\begin{equation}
\begin{split}
& \Gamma_{a \pm} = \frac{-\gamma_1 t}{2} \sqrt{rs} \Delta [1-i \gamma_1 \mathcal{W}^2 t](e^{\mathcal{W}t/2}),  \Gamma_{b \pm} =  \frac{ \gamma_1 t}{2} \sqrt{rs} \Delta [1-i \gamma_1 \mathcal{W'}^2 t] (e^{\mathcal{W'}t/2}),
\end{split}
\end{equation}
where $\mathcal{W} = \sqrt{rs} (\sqrt{rs} + \eta)$ and $\mathcal{W'} = \sqrt{rs} (\sqrt{rs} - \eta)$. From these expressions, one can derive the value of the squeezing parameter $\textbf{r}$ for $\eta = 0$, where $\Gamma_{a\pm} = \Gamma_{b\pm}$ is
\begin{equation}
\textbf{r}_a = \textbf{r}_b = \frac{ \gamma_1 t}{2} \sqrt{\mathcal{W}} \Delta [1-i \gamma_1 \mathcal{W}^2 t],
\end{equation}
for $\phi_a = \phi_b = \frac{\mathcal{W}t}{2}$.
\subsubsection{Entanglement Calculation}
Here, we are analyzing the evolution of entanglement by calculating the logarithmic negativity from the symplectic eigenvalues of a covariance matrix $\sigma$. For which we purposefully consider the two single-mode squeezed state $\xi$, defined as
\begin{equation}
    \ket{\xi} = \ket{\xi_a} \otimes \ket{\xi_b},
\end{equation}
and $\ket{\xi_a} = S(\xi_a) \ket{0}= \exp(\xi A^{\dagger 2} - \xi^* A^2) \ket{0}$ for $\xi = \textbf{r} e^{i \phi}$. We know that the squeezed vacuum state is referred to as a Gaussian state \cite{rai2010quantum}. Now, the corresponding covariance matrix is given by
\begin{equation}
\sigma=
\begin{pmatrix}
c & 0 & 0 & e\\
0 & d & e & 0\\
0 & e & c & 0\\
e & 0 & 0 & d
\end{pmatrix}
= 
\begin{pmatrix}
A & C\\
C^T & B
\end{pmatrix},
\end{equation}
where
\begin{equation}
\begin{split}
& c= \frac{1}{2} [\cosh(2 \textbf{r})+\sinh(2 \textbf{r}) \cos(2\phi)], d= \frac{1}{2} [\cosh(2 \textbf{r})-\sinh(2 \textbf{r}) \cos(2\phi)],e = \frac{1}{2} \sinh(2 \textbf{r}) \sin(2 \phi). 
\end{split}
\end{equation}
Since the entanglement is nothing but the logarithmic negativity of the lowest symplectic eigenvalue of a transpose covariance matrix given by
$E(\tilde{\sigma}) = max\{ 0,-\log (2\tilde{d_-})\}$ \cite{vidal2002computable} and the expression for the eigenvalues of a transpose covariance matrix are
\begin{equation}
\begin{split}
\tilde{d}_{\pm} & = \sqrt{\frac{\tilde{\Delta}(\sigma) \pm \sqrt{\tilde{\Delta}(\sigma)^2 - 4 Det(\sigma)}}{2}},
\end{split}
\end{equation}
where
\begin{equation}
 \tilde{\Delta}(\sigma) = Det(A) + Det(B) - 2 Det(C).
\end{equation}
\begin{figure}[ht]
    \centering
    \includegraphics[width=0.75\textwidth]{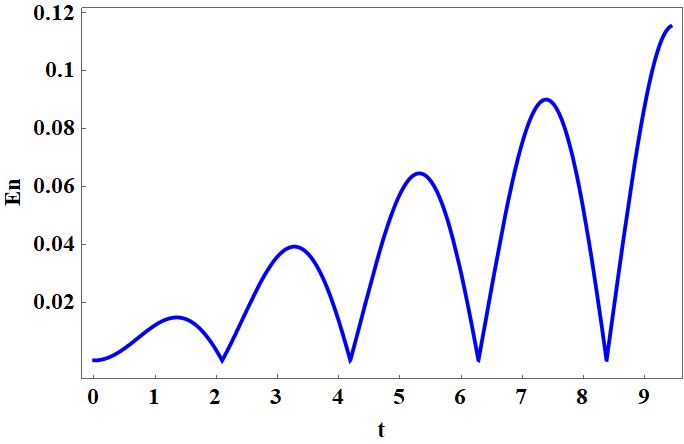}
    \caption{Time evolution of logarithmic negativity $E_n$ with respect to $t$. Here $\gamma_1$ = 0.1, $\mathcal{W}$ = 1.5 and $\Delta$ = 0.1.}
    \label{fig:eta3}
\end{figure}
From the expressions of Eq. (65), (69), and (70), one can easily verify that the eigenvalues are dependent on the squeezing parameter, which is a function of time. In Fig. \ref{fig:eta3}, we can see that the entanglement increases continuously with the increase of time. Initially, at $t=0$, $E_n=0$, indicating the state is separable. With increasing $t$ value, it oscillates between zero and non-zero values, which means the state becomes entangled and disentangled with growing time, which is initially separable.

\subsubsection{Quantum Mutual Information Calculation}
The expression for quantum mutual information is \cite{olivares2012quantum,slepian2003noiseless}:
\begin{equation}
I_M (\rho_{AB}) = S_V(\rho_A) + S_V(\rho_B) - S_V(\rho_{AB}),
\end{equation}
where $I_M$ denotes the quantum mutual information and $S_V(\rho) = - Tr[\rho \ln \rho]$, known as the von Neumann entropy. The above expression can be easily rewritten in terms of symplectic eigenvalues as:
\begin{equation}
 I_M (\rho_{AB}) = f(\sqrt{Det A}) + f(\sqrt{Det B}) - f(d_+) - f(d_-),
\end{equation}
where $d_{\pm}$ are the symplectic eigenvalues of the matrix $\sigma$ and  $f(x) = (x+\frac{1}{2}) \ln (x+\frac{1}{2}) - (x-\frac{1}{2}) \ln (x-\frac{1}{2})$. 

\begin{figure}[ht]
    \centering
    \includegraphics[width=0.75\textwidth]{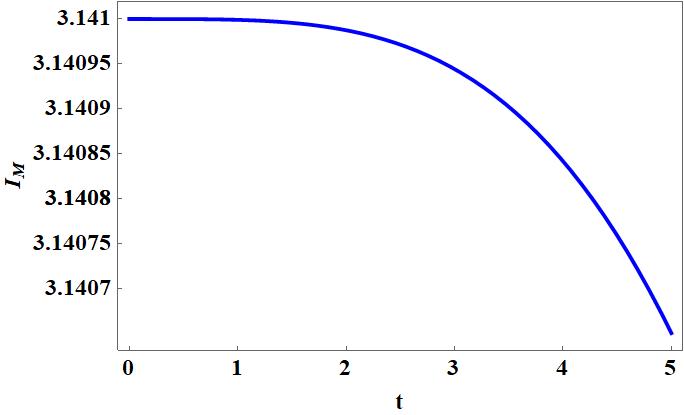}
    \caption{Time evolution of quantum mutual information with respect to $t$. Here $\gamma_1$ = 0.1, $\mathcal{W}$ = 0.1 and $\Delta$ = 0.1.}
    \label{fig:eta4}
\end{figure}

\section{Conclusion}
In our paper, we used the TFD technique to convert the master equation into a Schrödinger-like equation, whose Hamiltonian is similar to the Hamiltonian of the model described in the reference \cite{rai2010quantum}. Then, the application of the Hartree-Fock approximation linearizes the Hamiltonian. In our initial scenario, we consider the Hartree-Fock field and the external field to be similar, leading to $\eta=0$, and the non-tilden Hamiltonian is diagonalized by the Bogoliubov transformation, which can be explained by the evolution of operators. Similar method can be applicable to the tilden Hamiltonian. Additionally, by applying the disentanglement theorem and coherent states, we obtained the squeezing parameter and the wavefunction of the system. Since the Hamiltonian is diagonalized through the two-mode squeeze operator, after operating on the vacuum state, it is converted into a two-mode squeezed vacuum state, which leads to the construction of the covariance matrix. The symplectic eigenvalues of the transpose of the covariance matrix, associated with logarithmic negativity for a two-mode Gaussian state, provide a suitable method to study the entanglement. From Fig. \ref{fig:eta1}, it can be seen that the state is separable at $t=0$, and as the value of $ t$ increases, the state oscillates between zero and non-zero values, exhibiting both entangled and disentangled behavior. Lastly, we successfully plotted the quantum mutual information graph with time $t$ by using the von Neumann entropy in Fig. \ref{fig:eta2}, followed by symplectic eigenvalues of the covariance matrix. 

In the second case, we took the Hartree-Fock field and the external field are not similar, for $\eta \ne 0$. As the nontilden Hamiltonian is associated with $su(1,1)$ symmetry, the Hamiltonian can be rewritten in terms of $su(1,1)$ operators, followed by the Bogoliubov transformation. Similarly, the tilden Hamiltonian can also be expressed in terms of $su(1,1)$ operators. The usage of the disentanglement theorem and the application of coherent states in the vacuum give rise to the squeezing parameter and the wave function. Here, the Hamiltonian is diagonalized by the unitary transformation associated with the two single-mode squeezed states, which gives rise to the covariance matrix. The entanglement of the state was analyzed through the logarithmic negativity, calculated by the eigenvalues of the transpose of the covariance matrix. In Fig. \ref{fig:eta3}, one can conclude that the state is separable initially and with growing time, the state oscillates between zero to non-zero values, showing the entangled and disentangled nature. Lastly, in Fig. \ref{fig:eta4} we analyzed the quantum mutual information through the symplectic eigenvalues of the constructed covariance matrix and the usage of the von Neumann entropy. These methods applied for solving the master equation will find many applications in quantum information theory, quantum optics, building quantum computers, quantum channels, and open quantum systems. 

\section{Appendices}
\subsection{Appendix A: Basics of Thermofield Dynamics}
A brief description of TFD is given below. 
The dissipative term in any master equations makes it difficult
to apply the usual Schrödinger equation techniques with pure states to mixed states. The thermo field dynamics (TFD) provides such a formalism. 
In TFD, the mixed state averages are expressed as scalar products, and the dynamics is given in terms of a Schrödinger-like equation. A density operator $\rho=\vert N\rangle\langle N\vert$ corresponding to a Fock state $\vert N\rangle$ in the Hilbert space ${\cal H}$ is viewed in TFD as a vector $\rho=\vert N, \tilde{N} \rangle$ in an extended Hilbert space ${\cal H}\otimes {\cal H}^*$. The central idea in TFD is to construct a density operator $\vert \rho^\alpha\rangle,  1/2  \le \alpha \le 1$ as a vector in the extended Hilbert space ${\cal H}\otimes {\cal H}^*$. 

Here the averages of operators with respect to $\rho$ reduces to a scalar product:
\begin{equation}
\tag{A.1}
\langle A\rangle = Tr [A\rho] = \langle\rho^{1-\alpha}\vert A\vert \rho^\alpha\rangle,
\end{equation}
where $\vert \rho^\alpha\rangle$ is given by
\begin{equation}
\tag{A.2}
\vert \rho^\alpha\rangle = \hat{\rho}^\alpha\vert I\rangle\;, \text{with},\; \hat{\rho}^\alpha = \rho^\alpha \otimes I,
\end{equation}
where $\vert I\rangle$ is the resolution of the
identity
\begin{equation}
\tag{A.3}
\vert I\rangle=\sum \vert n\rangle\langle n\vert = \sum \vert n\rangle\otimes\vert \tilde{n}\rangle \equiv \sum \vert n,\tilde{n}\rangle,
\end{equation}
in terms of a complete orthonormal basis  $\{\vert n\rangle\}_{n=0}^\infty$  in  ${\cal  H}$.
The state vector $\vert I\rangle$ takes a normalized vector to another
normalized vector in the extended Hilbert space ${\cal H}\otimes {\cal 
H}^*$. The matrix $A(a,a^\dagger)$  acts like $A \otimes I$. 

(It may be noted that for any density operator the states $\vert \rho^\alpha\rangle,  1/2  \le \alpha \le 1$ have a finite norm in the extended Hilbert space ${\cal H}\otimes {\cal H}^*$. This is not in general true for the state $\vert \rho^{1-\alpha}\rangle,  1/2  \le \alpha \le 1$, which includes $\vert I\rangle$. These states are regarded as formal but extremely useful constructs.)\\
The creation and the annihilation operators $a^\dagger, \tilde{a}^\dagger,  a$,  and  $\tilde{a}$ are introduced as follows
\begin{equation}
\tag{A.4}
\begin{split}
a\vert n,\tilde{m}\rangle & = \sqrt{n} \vert n-1,\tilde{m}\rangle,\nonumber\\ 
a^\dagger\vert n,\tilde{m}\rangle &= \sqrt{n+1} \vert n+1,\tilde{m}\rangle,   \end{split}
\end{equation}
\begin{equation}
\tag{A.5}
\begin{split}
\tilde{a}\vert n,\tilde{m}\rangle &= 
\sqrt{m} \vert n,\tilde{m}-1\rangle,\nonumber\\ 
\tilde{a}^\dagger\vert n,\tilde{m}\rangle &= \sqrt{m+1} \vert n,\tilde{m}+1\rangle.    
\end{split}
\end{equation}
The operators $a$ and $a^\dagger$  commute  with  $\tilde{a}$  and $\tilde{a}^\dagger$.  It is clear from the above that $a$ acts on the vector space $\cal{H}$ and $\tilde{a}$ acts on the vector space $\cal{H^*}$. From the expression for $\vert I\rangle$ in terms of the number states
\begin{equation}
\tag{A.6}
\vert I\rangle = \sum_n \vert n,\tilde{n}\rangle,
\end{equation}
it follows that
\begin{equation}
\tag{A.7}
a\vert I\rangle=\tilde{a}^\dagger \vert I\rangle,\; a^\dagger\vert I\rangle = 
\tilde{a}\vert I\rangle,
\end{equation}
and hence for any operator $A$ (written in terms of $a$ $a^\dagger$ and their complex conjugates), one has
\begin{equation}
\tag{A.8}
A\vert I\rangle = \tilde{A}^\dagger \vert I\rangle,
\end{equation}
where $\tilde{A}$ is obtained from $A$ by making the  replacements 
tilde  conjugation  rules   $a\to   \tilde{a},   a^\dagger   \to 
\tilde{a}^\dagger,  \alpha\to  \alpha^*$. 
An immediate consequence of this is that  the state $\vert \rho^\alpha\rangle$
which remains unchanged under the replacements $a\to   \tilde{a}$,  
$a^\dagger  \to \tilde{a}^\dagger,$  and c number $\to$ complex conjugates C by applying the  the hermiticity property of $\rho$
i.e. $\rho^\dagger=\rho$. 
The tildian property reflects the hermiticity property of the density operator.\\
The evolution of a conservative system in terms of $\rho^\alpha$ is given by the von Neumann equation
\begin{equation}
\tag{A.9}
 \frac{\partial}{\partial t}\rho^\alpha(t)=\frac{-i}{\hbar}[H,\rho^\alpha],
\end{equation}
by applying $\vert I\rangle$ from the right and one gets
\begin{equation}
\tag{A.10}
i \frac{\partial}{\partial t} \rho^\alpha \ket{I} = H \rho^\alpha \ket{I} - \rho^\alpha H \ket{I}.
\end{equation}
In order to obtain the equation for $\ket{\rho^\alpha} \equiv \hat{\rho}^\alpha \ket{I}$, $\rho^\alpha$ present in the second term of the R.H.S. of Eq. (A.10) must move next to $\ket{I}$ by using the relations in Eq. (A.7) and the Hermiticity of $H$. This operation gives rise to the following equation:
\begin{equation}
\tag{A.11}
\frac{\partial}{\partial t}\vert \rho^\alpha(t)\rangle = -i\hat{H}\vert \rho^\alpha\rangle,
\end{equation}
where
\begin{equation}
\tag{A.12}
 \hat{H}=(H-\tilde{H}).
\end{equation}
In TFD, one can derive a Schrödinger-like equation for any state $\vert\rho^\alpha\rangle$
with an arbitrary value of $\alpha$.   
For dissipative systems, the evolution equation is given by master equation 
\begin{equation}
\tag{A.13}
 \frac{\partial}{\partial t}\rho(t)=\frac{-i}{\hbar}(H\rho-\rho H)+L\rho,
\end{equation}
where $L$ is the Liouville term. The non-equilibrium thermofield dynamics is developed and analysed in $\alpha=1$ representation. Hence, from now on, we work in, $\alpha=1$ representation \cite{shanta1996eigenstates,shanta1996operator,chaturvedi1999quantum}. In this representation, for any hermitian operator $A$, one has 
\begin{equation}
\tag{A.14}
\langle A\rangle = \langle I\vert A\vert \rho\rangle = \langle A\vert\rho\rangle= Tr(A\rho).
\end{equation} 
By applying $\vert I\rangle$ to Eq. (A.13) from the right, one goes over to TFD, and the master equation is given by Eq. (9). 
With $-i\hat{H}$ being a tildian, and thus the problem of solving the master equation is reduced to solving a Schrödinger-like equation, namely Eq. (9). 

Historically, the thermo field dynamics was developed in $\alpha=\frac{1}{2}$ representation. In this representation  $ \vert \rho_0^\frac{1}{2}\rangle$ is related to the $\vert 0,0\rangle$ by a unitary transformation,
which is nothing but the Caves-Schumaker state, for details ref \cite{leplae1974derivation,takahashi1996thermo,ojima1981gauge,umezawa1982thermo,shanta1996operator,chaturvedi1990thermal}.

\subsection{Appendix B: Hartree-Fock approximation}
The Hartree-Fock approximation plays a crucial role in solving the nonlinear master equation of an open quantum system. The expression of the Hamiltonian in Eq. (17) is a nonlinear one with the presence of fourth-order operator terms, which can not be easily resolved through conventional methods. The Hartree-Fock approach offers an elegant solution by treating these complicated multi-operator expressions as products of simpler two-operator averages, effectively transforming an intractable nonlinear problem into a manageable linear one \cite{chaturvedi1990use,shiv2014entanglement}.\\ 
So applying the Hartree-Fock approximation in each term of Eq. (17), we have 
\begin{equation}
\tag{B.1}
\begin{split}
& 2a^{\dagger} b \tilde{b} \tilde{a}^{\dagger} = -2ba^{\dagger} \tilde{b} \tilde{a}^{\dagger} = - \langle b a^{\dagger} \rangle \tilde{b} \tilde{a}^{\dagger} - b a^{\dagger} \langle \tilde{b} \tilde{a}^{\dagger} \rangle,\\
& b^{\dagger} a a^{\dagger} b = a b^{\dagger} b a^{\dagger} = \frac{\langle ab^{\dagger} \rangle}{2} ba^{\dagger} + ab^{\dagger} \frac{\langle ba^{\dagger} \rangle}{2},\\
& \tilde{a} \tilde{b}^{\dagger} \tilde{b} \tilde{a}^{\dagger} = - \tilde{b} \tilde{b}^{\dagger} + \tilde{a} \tilde{a}^{\dagger} + \frac{ \langle\tilde{b} \tilde{a}^{\dagger} \rangle}{2} \tilde{a} \tilde{b}^{\dagger} + \tilde{b} \tilde{a}^{\dagger} \frac{\langle\tilde{a} \tilde{b}^{\dagger} \rangle}{2}.
\end{split}
\end{equation}
And taking, $ \langle ba^{\dagger} \rangle = \Delta_1 , \langle ab^{\dagger} \rangle = \Delta^*_1 ,
\langle \tilde{b} \tilde{a}^{\dagger} \rangle = \tilde{\Delta}_1(t)$, $\langle \tilde{a} \tilde{b}^{\dagger} \rangle = \tilde{\Delta}_1^*$ and $\Delta_1(t) = \Delta^*_1$, $\tilde{\Delta}_1 = \tilde{\Delta}^*_1 $, $-1\hat{H}_1$,  Eq. (17) becomes
\begin{equation}
\tag{B.2}
\begin{split}
  -i \hat{H_1} = & \frac{1}{i \hbar} \Bigl[\frac{\epsilon_0}{2}(a^{\dagger} a - \tilde{a} \tilde{a}^{\dagger}) + \epsilon_1 (b^{\dagger} b - \tilde{b} \tilde{b}^{\dagger})
+ g_1 \biggl\{ (a^{\dagger} b - \tilde{a} \tilde{b}^{\dagger}) E_1 + (b^{\dagger} a - \tilde{b} \tilde{a}^{\dagger}) E_1^* \biggr\} \Bigr] \\
& + \frac{\gamma_1}{2} \biggl[ -a^{\dagger}a + b^{\dagger}b - (\Delta_1 + \tilde{\Delta}_1 (ba^{\dagger} + ab^{\dagger}) -\tilde{a} \tilde{a}^{\dagger} + \tilde{b} \tilde{b}^{\dagger} - (\Delta_1 + \tilde{\Delta}_1 (\tilde{b} \tilde{a}^{\dagger} + \tilde{a} \tilde{b}^{\dagger}) \biggr].
\end{split}
\end{equation}

\subsection{Appendix C: Bogoliubov transformation (for \texorpdfstring{$\eta=0$}{eta=0})}

From Eq. (24), the Hamiltonian is given by
\begin{equation}
\tag{C.1}
-iH_1 = rs(a^\dagger a+b^\dagger b) - \frac{\gamma_1\Delta}{2}\sqrt{rs}(ab + a^\dagger b^\dagger).
\end{equation}
With the application of the following Bogoliubov transformation \cite{shiv2014entanglement,chalker2013quantum,birol2007bogoliubov,emary2001bogoliubov}
\begin{equation}
\tag{C.2}
\begin{split}
& A = \mu a + \nu^* b^{\dagger}, A^{\dagger}  = \mu^* a^{\dagger} + \nu b,\\
& B = \mu b + \nu^* a^{\dagger}, B^{\dagger}  = \mu^* b^{\dagger} + \nu a,
\end{split}
\end{equation}
with $\mu= \mu^* = \cosh{\textbf{r}}$ and $\nu^* = \sinh{\textbf{r}} e^{i \phi}, \nu = \sinh{\textbf{r}} e^{-i \phi}$we can diagonalize the above Hamiltonian. Now, by writing the Hamiltonian in a convenient form
\begin{align}
\tag{C.3}
     -iH_1 &=
\begin{pmatrix}
   a^\dagger & b
\end{pmatrix}
\begin{pmatrix}
m & n\\
n & m
\end{pmatrix}
\begin{pmatrix}
   a \\
   b^\dagger
\end{pmatrix},
\end{align}
where $m = rs $ \& $n = - \frac{\gamma_1\Delta}{2}\sqrt{rs}$. By defining the coefficient matrix in Eq.(C.1) with
\begin{equation}
\tag{C.4}
\begin{pmatrix}
m & n\\
n & m
\end{pmatrix}=
\begin{pmatrix}
\cosh \mathbf{r} & \sinh \mathbf{r}\\
\sinh \mathbf{r} & \cosh \mathbf{r}
\end{pmatrix},
\end{equation}
and with a preferable normalization $\sqrt{|m|^2 -|n|^2}$, the Bogoliubov coefficients can be read off as
\begin{align}
\tag{C.5}
     &\mu = \cosh{\mathbf{r}} = \mathcal{W},
    \nu = \sinh{\mathbf{r}} = - \frac{\gamma_1\Delta}{2}\sqrt{\mathcal{W}}, 
\end{align}
where $\mathcal{W} = rs$.

\subsection{Appendix D: \texorpdfstring{$\Delta$}{Delta} Calculation: Using Bogoliubov transformation (for \texorpdfstring{$\eta=0$}{eta=0})}
Considering the following field variables appear in Eq. (24)
\begin{equation*}
    \Delta = \langle ab \rangle.
\end{equation*}
The initial state is the thermal state in usual notation, given by
\begin{equation*}
    \rho (t) = \sum_{n,m = 0}^{\infty} \frac{\Bar{n}_1^n}{{(\Bar{n}_1 + 1)}^{n+1}} \frac{\Bar{n}_2^m}{{(\Bar{n}_2 + 1)}^{m+1}} e^{-i \int{dt H}} \ket{n,m} \bra{n,m} e^{i \int{dt H}},
\end{equation*}
where $H$ is equal to $-iH_1$ in Eq. (24). So, we have
\begin{equation*}
\begin{split}
\langle ab \rangle = &  \langle I|ab|\rho(t) \rangle = \Tr(ab \rho(t)),\\
= & \sum_{n,m = 0}^{\infty} \frac{\Bar{n}_1^n}{{(\Bar{n}_1 + 1)}^{n+1}} \frac{\Bar{n}_2^m}{{(\Bar{n}_2 + 1)}^{m+1}} \Tr(e^{-i \int{dt H}} \ket{n,m} \bra{n,m} e^{i \int{dt H}}),\\
= & \sum_{n,m = 0}^{\infty} \frac{\Bar{n}_1^n}{{(\Bar{n}_1 + 1)}^{n+1}} \frac{\Bar{n}_2^m}{{(\Bar{n}_2 + 1)}^{m+1}} \sum_{p,q} \bra{p,q}e^{-i \int{dt H}} \ket{n,m} \bra{n,m} e^{i \int{dt H}} \ket{p,q},\\
= & \sum_{n,m = 0}^{\infty} \frac{\Bar{n}_1^n}{{(\Bar{n}_1 + 1)}^{n+1}} \frac{\Bar{n}_2^m}{{(\Bar{n}_2 + 1)}^{m+1}} \sum_{p,q} \bra{n,m}e^{i \int{dt H}} \ket{p,q} \bra{p,q} e^{-i \int{dt H}} \ket{n,m},\\
= & \sum_{n,m = 0}^{\infty} \frac{\Bar{n}_1^n}{{(\Bar{n}_1 + 1)}^{n+1}} \frac{\Bar{n}_2^m}{{(\Bar{n}_2 + 1)}^{m+1}} \bra{n,m}e^{i \int{dt H}} \biggl(\sum_{p,q} \ket{p,q} \bra{p,q} \biggr) e^{- i \int{dt H}} \ket{n,m},\\
= & \bra{0,0} e^{i \int{dt H}} ab e^{-i \int{dt H}} \ket{0,0}\\ 
+ & \sum_{n,m = 1}^{\infty} \frac{\Bar{n}_1^n}{{(\Bar{n}_1 + 1)}^{n+1}} \frac{\Bar{n}_2^m}{{(\Bar{n}_2 + 1)}^{m+1}} \bra{n,m}e^{i \int{dt H}} ab  e^{-i \int{dt H}} \ket{n,m}.
\end{split}
  \end{equation*}
Considering
\begin{equation}
\tag{D.1}
\begin{split}
e^{i \int{dt H}} ab  e^{-i \int{dt H}} = & (1 + i \int{dt H}) ab (1 - i \int{dt H}) = ab + i\int{dt [H, ab]},
\end{split}
\end{equation}
and taking the average value of $\langle ab \rangle$, given as
\begin{equation}
\tag{D.2}
\begin{split}
\langle ab \rangle = & \bra{0,0} e^{i \int{dt H}} ab e^{-i \int{dt H}} \ket{0,0} + \sum_{n,m = 0}^{\infty} \frac{\Bar{n}_1^n}{{(\Bar{n}_1 + 1)}^{n+1}} \frac{\Bar{n}_2^m}{{(\Bar{n}_2 + 1)}^{m+1}} \bra{n,m}e^{i \int{dt H}} ab  e^{-i \int{dt H}} \ket{n,m},
\end{split}
\end{equation}
we have calculated the $\Delta$ value, given as
\begin{equation}
\tag{D.3}
\begin{split} 
\langle ab \rangle = &\bra{0,0} ab \ket{0,0} + i\int{dt \bra{0,0}[H, ab] \ket{0,0}}\\
+ & \sum_{n,m = 0}^{\infty} \frac{\Bar{n}_1^n}{{(\Bar{n}_1 + 1)}^{n+1}} \frac{\Bar{n}_2^m}{{(\Bar{n}_2 + 1)}^{m+1}} \biggl( \bra{n,m} ab \ket{n,m} + i \int{dt \bra{n,m} [H, ab] \ket{n,m}} \Biggr).
\end{split}
\end{equation}
Using the $su(1,1)$ operators in the above equation and applying the inverse squeezing transformations of Eq. (25), one can get
\begin{align*}
\langle ab \rangle & = c \Delta - \mu^*\nu^* + i \int {dt}  \left[rs \mu^*\nu^* + \frac{\gamma_1\Delta}{2}\sqrt{rs} \left( 1 + 2 \lvert\nu\rvert^2 \right)\right]
    \\
    &- \sum_{m,n=1}^{\infty} \frac{\bar{n}_1^n}{(\bar{n}_1+1)^{n+1}}\frac{\bar{n}_2^m}{(\bar{n}_2+1)^{m+1}}\mu^*\nu^*(n+m+1)+\sum_{m,n=1}^{\infty} \frac{\bar{n}_1^n}{(\bar{n}_1+1)^{n+1}}\frac{\bar{n}_2^m}{(\bar{n}_2+1)^{m+1}} \times
    \\
    &i\int {dt}\left[rs\mu^*\nu^*(n+m+1) +
    \frac{\gamma_1\Delta}{2}\sqrt{rs}(1 +
    (\lvert\mu\rvert^2 + \lvert\nu\rvert^2)(m+n) +
    2\lvert\nu\rvert^2)\right]
    \\
    &=  c \Delta - \mu^*\nu^* +i \int {dt}\, \mathcal{W} \mu^*\nu^* + i \int {dt}\, \frac{\gamma_1\Delta}{2}\sqrt{\mathcal{W}} \left( 1 + 2 \lvert\nu\rvert^2 \right)
    \\
    & + \sum_{m,n=1}^{\infty} \frac{\bar{n}_1^n}{(\bar{n}_1+1)^{n+1}}\frac{\bar{n}_2^m}{(\bar{n}_2+1)^{m+1}} \Big[- \mu^*\nu^*(n+m+1) + i\int {dt}\,\mathcal{W}\mu^*\nu^*(n+m+1)
    \\
    &+ i\int {dt}\, \frac{\gamma_1\Delta}{2}\sqrt{\mathcal{W}}(1 +
    (\lvert\mu\rvert^2 + \lvert\nu\rvert^2)(m+n) +
    2\lvert\nu\rvert^2)\Big],
\end{align*}
where $c\Delta$ is an integration constant at $t=0$. Putting the values of $\mu$ and $\nu$ from the above Bogoliubov transformation from Eq. (C.5) and since the $\Delta$ calculation is a self-consistency calculation, substituting the $\Delta$ value again in the above calculation, we have
\begin{equation}
\tag{D.4}
\begin{split}
    \Delta &= \langle ab \rangle\\
    &= c\Delta +  \frac{\gamma_1\Delta}{2}\mathcal{W}\sqrt{\mathcal{W}} \left(1 - i \mathcal{W} t + i \frac{\gamma_1t\sqrt{\mathcal{W}}}{2}\right)+ i\frac{\gamma_1^2\mathcal{W}^2}{2} \int{dt}\,\Delta\sinh^2{\mathbf{r}}\\
    &+ \sum_{m,n=1}^{\infty} \frac{\bar{n}_1^n}{(\bar{n}_1+1)^{n+1}}\frac{\bar{n}_2^m}{(\bar{n}_2+1)^{m+1}}\Big[-\frac{1}{2}\sinh{2 \mathbf{r}}(m+n+1) + \frac{i \mathcal{W}}{2}\int {dt}\,\sinh{2 \mathbf{r}}(m+n+1)\\
    &+ i\int {dt}\, \frac{\gamma_1\Delta}{2}\sqrt{\mathcal{W}}\left( 1 + \cosh{2 \mathbf{r}} (m+n) + 2 \cosh^2{\mathbf{r}} \right) \Big].
\end{split}
\end{equation}
Here, $\Delta$ is calculated for the short time \cite{shiv2014entanglement}.

\subsection{Appendix E: Calculation of Bogoliubov transformation (for \texorpdfstring{$\eta \neq 0$}{eta ne 0})}
From Eq. (48), the Hamiltonian is given by
\begin{equation}
\tag{E.1}
\begin{split}
    -iH'_1 = & \sqrt{rs} (\sqrt{rs} + \eta) (A^{\dagger} A +  A A^{\dagger}) - \frac{\gamma_1 \Delta}{2} \sqrt{rs} (A^2 + A^{\dagger 2})\\
    & + \sqrt{rs} (\sqrt{rs} - \eta) (B^{\dagger} B +  B B^{\dagger}) + \frac{\gamma_1 \Delta}{2} \sqrt{rs} (B^2 + B^{\dagger 2}) - 2rs.
\end{split}
\end{equation}
With the application of the following Bogoliubov transformation \cite{shiv2014entanglement,chalker2013quantum,birol2007bogoliubov,emary2001bogoliubov}
\begin{equation}
\tag{E.2}
\begin{split}
& X = \mu A + \nu^* A^{\dagger}, X^{\dagger}  = \mu^* A^{\dagger} + \nu A,\\
& Y = \mu B + \nu^* B^{\dagger}, Y^{\dagger}  = \mu^* B^{\dagger} + \nu B,
\end{split}
\end{equation}
with $\mu= \mu^* = \cosh{\textbf{r}}$ and $\nu^* = \sinh{\textbf{r}} e^{i \phi}, \nu = \sinh{\textbf{r}} e^{-i \phi}$we can diagonalize the above Hamiltonian. Now, by writing the Hamiltonian in a convenient form
\begin{align}
\tag{E.3}
     -iH_{1A} &=
\begin{pmatrix}
   A^\dagger & A
\end{pmatrix}
\begin{pmatrix}
m & n\\
n & m
\end{pmatrix}
\begin{pmatrix}
   A \\
   A^\dagger
\end{pmatrix},
-iH_{1B} =
\begin{pmatrix}
   B^\dagger & B
\end{pmatrix}
\begin{pmatrix}
m & n\\
n & m
\end{pmatrix}
\begin{pmatrix}
   B \\
   B^\dagger
\end{pmatrix},
\end{align}
where $m = \sqrt{rs} (\sqrt{rs} + \eta) = \sqrt{rs} (\sqrt{rs} - \eta)$ and $ n = \frac{\gamma_1 \Delta}{2} \sqrt{rs}$. By defining the coefficient matrix in Eq.(E.1) with
\begin{equation}
\tag{E.4}
\begin{pmatrix}
m & n\\
n & m
\end{pmatrix}=
\begin{pmatrix}
\cosh \mathbf{r} & \sinh \mathbf{r}\\
\sinh \mathbf{r} & \cosh \mathbf{r}
\end{pmatrix},
\end{equation}
and with a preferable normalization $\sqrt{|m|^2 -|n|^2}$, the Bogoliubov coefficients can be read off as
\begin{equation}
\tag{E.5}
\begin{split}
& \mu_A = \cosh (\mathbf{r}) = \frac{\sqrt{rs} (\sqrt{rs} + \eta)}{\sqrt{[\sqrt{rs} (\sqrt{rs} + \eta)]^2 - [\frac{\gamma_1 \Delta}{2} \sqrt{rs}]^2}}, \hspace{1mm} \mu_B = \cosh (\mathbf{r}) = \frac{\sqrt{rs} (\sqrt{rs} - \eta)}{\sqrt{[\sqrt{rs} (\sqrt{rs} - \eta)]^2 - [\frac{\gamma_1 \Delta}{2} \sqrt{rs}]^2}},\\
& \nu_A = \sinh(\mathbf{r}) = \frac{\frac{\gamma_1 \Delta}{2} \sqrt{rs}}{\sqrt{[\sqrt{rs} (\sqrt{rs} + \eta)]^2 - [\frac{\gamma_1 \Delta}{2} \sqrt{rs}]^2}}, \hspace{1mm} \nu_B = \sinh(\mathbf{r}) = \frac{\frac{\gamma_1 \Delta}{2} \sqrt{rs}}{\sqrt{[\sqrt{rs} (\sqrt{rs} - \eta)]^2 - [\frac{\gamma_1 \Delta}{2} \sqrt{rs}]^2}},
\end{split}
\end{equation}
with
\begin{equation}
\tag{E.6}
    |\mu|^2 - |\nu|^2 = 1.
\end{equation}

\subsection{Appendix F: \texorpdfstring{$\Delta$}{Delta} Calculation (for \texorpdfstring{$\eta \neq 0$}{eta neq 0})}
From Eq. (48), the Hamiltonian is given by
\begin{equation}
\tag{F.1}
\begin{split}
    -iH'_1 = & \sqrt{rs} (\sqrt{rs} + \eta) (A^{\dagger} A +  A A^{\dagger}) - \frac{\gamma_1 \Delta}{2} \sqrt{rs} (A^2 + A^{\dagger 2})\\
    & + \sqrt{rs} (\sqrt{rs} - \eta) (B^{\dagger} B +  B B^{\dagger}) + \frac{\gamma_1 \Delta}{2} \sqrt{rs} (B^2 + B^{\dagger 2}) - 2rs.
\end{split}
\end{equation}
By doing the calculation, we have
\begin{equation}
\tag{F.2}
\begin{split}
\Delta(t) = \langle ab \rangle = & c \Delta -i \int{dt 2 \gamma_1 \Delta (|\mu|^2 + |\nu|^2)}\\
- & \sum_{n,m = 0}^{\infty} \frac{\Bar{n}_1^n}{{(\Bar{n}_1 + 1)}^{n+1}} \frac{\Bar{n}_2^m}{{(\Bar{n}_2 + 1)}^{m+1}} i \int{dt \hspace{1mm} 2\gamma_1 \Delta \cosh{(2 \mathbf{r})} [m + n + 1]}. 
\end{split}
\end{equation}
Putting the $\mu$ and $\nu$ from Eq. (E.5) for $\eta=0$, we got
\begin{equation}
\tag{F.3}
\begin{split}
\Delta = \langle ab \rangle = & c\Delta -i \int{dt 2 \gamma_1 \Delta (|\mu|^2 + |\nu|^2)}\\
- & \sum_{n,m = 1}^{\infty} \frac{\Bar{n}_1^n}{{(\Bar{n}_1 + 1)}^{n+1}} \frac{\Bar{n}_2^m}{{(\Bar{n}_2 + 1)}^{m+1}} i \int{dt \hspace{1mm} 2\gamma_1 \Delta \cosh{(2 \mathbf{r})} [m + n + 1]},\\
& = c \Delta - i \int dt 2 \gamma_1 \Delta \biggl[ \mathcal{W}^2 + \biggl(\frac{\gamma_1 \Delta}{2} \sqrt{rs} \biggr)^2 \biggr] \\
- & \sum_{n,m = 1}^{\infty} \frac{\Bar{n}_1^n}{{(\Bar{n}_1 + 1)}^{n+1}} \frac{\Bar{n}_2^m}{{(\Bar{n}_2 + 1)}^{m+1}} i \int{dt \hspace{1mm} 2\gamma_1 \Delta \cosh{(2 \mathbf{r})} [m + n + 1]},\\
& = c \Delta -2i \gamma_1 \Delta \mathcal{W}^2 t - i \frac{\gamma_1^3}{2} \mathcal{W} [\Delta]^3 t\\
- & \sum_{n,m = 1}^{\infty} \frac{\Bar{n}_1^n}{{(\Bar{n}_1 + 1)}^{n+1}} \frac{\Bar{n}_2^m}{{(\Bar{n}_2 + 1)}^{m+1}} i \int{dt \hspace{1mm} 2\gamma_1 \Delta \cosh{(2 \mathbf{r})} [m + n + 1]},\\
\end{split}
\end{equation}
where c = integration constant at $t=0$ and $\mathcal{W} = \mathcal{W'}$ for $\eta =0$. Here, $\Delta$ is calculated for the short time \cite{shiv2014entanglement}.

\subsection{Appendix G: \texorpdfstring{$\Delta_1$}{Delta1} Calculation (for \texorpdfstring{$\eta \neq 0$}{eta neq 0})}

From Eq. (18), we have
\begin{equation}
    \tag{G.1}
      \langle a^{\dagger} b \rangle = \Delta_1,
\end{equation}
and
\begin{equation}
\tag{G.2}
\begin{split} 
\langle a^{\dagger}b \rangle = &\bra{0,0} a^{\dagger}b \ket{0,0} + i\int{dt \bra{0,0}[H, a^{\dagger}b] \ket{0,0}}\\
+ & \sum_{n,m = 0}^{\infty} \frac{\Bar{n}_1^n}{{(\Bar{n}_1 + 1)}^{n+1}} \frac{\Bar{n}_2^m}{{(\Bar{n}_2 + 1)}^{m+1}} \biggl( \bra{n,m} a^{\dagger}b \ket{n,m} + i \int{dt \bra{n,m} [H, a^{\dagger}b] \ket{n,m}} \Biggr).
\end{split}
\end{equation}
Following the above methodology, $\Delta_1$ can be expressed as
\begin{equation}
\tag{G.3}
\begin{split}
\Delta_1 = & \langle a^{\dagger}b \rangle = c\Delta_1 +2 \cosh{(2 \mathbf{r})} \\
+ & 4i \int{dt \Biggl[ rs (-\sinh{(2 \mathbf{r})} + \frac{\gamma_1 \Delta(t) \sqrt{rs}}{2} \cosh{(2 \mathbf{r})} \Biggr] \cosh{(\mathbf{r})} \sinh{(\mathbf{r})}}\\
+ & \sum_{n,m = 0}^{\infty} \frac{\Bar{n}_1^n}{{(\Bar{n}_1 + 1)}^{n+1}} \frac{\Bar{n}_2^m}{{(\Bar{n}_2 + 1)}^{m+1}} \cosh{(2 \mathbf{r})} (n-m)\\
+ & \sum_{n,m = 0}^{\infty} \frac{\Bar{n}_1^n}{{(\Bar{n}_1 + 1)}^{n+1}} \frac{\Bar{n}_2^m}{{(\Bar{n}_2 + 1)}^{m+1}}\\
& 4i \int{dt \Biggl[-rs \sinh{(2 \mathbf{r})} + \frac{\gamma_1 \Delta(t) \sqrt{rs}}{2} \cosh{(2 \mathbf{r})} \Biggr] \cosh(\mathbf{r}) \sinh(\mathbf{r}) (2m + 1)},
\end{split}
\end{equation}

where c = integration constant at $t=0$ and $ \cosh (\mathbf{r}) = \frac{rs }{\sqrt{[rs ]^2 - [\frac{\gamma_1 \Delta}{2} \sqrt{rs}]^2}}$, $\sinh(\mathbf{r}) = \frac{\frac{\gamma_1 \Delta}{2} \sqrt{rs}}{\sqrt{[rs]^2 - [\frac{\gamma_1 \Delta}{2} \sqrt{rs}]^2}}$ for $\eta =0$ (from Eq. (E.5)). Here, $\Delta_1$ is calculated for the short time \cite{shiv2014entanglement}.

\cleardoublepage
\addcontentsline{toc}{section}{BIBLIOGRAPHY}
\bibpunct{(}{)}{;}{n}{,}{,}
\bibliographystyle{unsrt}
\bibliography{reference}
\end{document}